\pdfoutput=1
\documentclass[aps,prl,twocolumn,showpacs]{revtex4}

\usepackage{graphicx}
\usepackage{textcomp}
\usepackage{amssymb}\usepackage{amsmath} 

\begin{document}

\title{Dynamical Coupling between a Bose-Einstein Condensate and a Cavity Optical Lattice}

\author{Stephan Ritter$^{1,2}$, Ferdinand Brennecke$^1$, Kristian Baumann$^1$, Tobias Donner$^{1,3}$, Christine Guerlin$^1$, Tilman Esslinger$^1$
}                     

\email{esslinger@phys.ethz.ch}

\affiliation{$^1$Institute for Quantum Electronics, ETH Z\"urich, CH-8093
Z\"urich, Switzerland\\
$^2$Max-Planck-Institut f\"ur Quantenoptik, 85748 Garching, Germany\\
$^3$JILA, University of Colorado and National Institute of Standards and Technology, Boulder, CO 80309, USA}

\date{Revised version: 11 February 2009}


\begin{abstract} 
A Bose-Einstein condensate is dispersively coupled to a single mode of an ultra-high finesse optical cavity. The system is governed by strong interactions between the atomic motion and the light field even at the level of single quanta. While coherently pumping the cavity mode the condensate is subject to the cavity optical lattice potential whose depth depends nonlinearly on the atomic density distribution. We observe optical bistability already below the single photon level and strong back-action dynamics which tunes the coupled system periodically out of resonance.
\end{abstract}
\maketitle
\bigskip
\section{Introduction}
\label{section:introduction}

The coherent interaction between matter and a single mode of light is a fundamental theme in cavity quantum electrodynamics \cite{haroche2006}. Experiments have been realized both in the microwave and the optical domain, with the cavity field being coupled to Rydberg atoms \cite{raimond2001,walther2002}, neutral atoms \cite{kimble1998,maunz2005a}, ions \cite{mundt2002} or artificial atoms like superconducting qubits \cite{wallraff2004}. The energy spectrum of these systems is characterized by an avoided crossing between the atomic and cavity excitation branches. Far detuned from the atomic resonance the dispersive regime is realized. The atom-light coupling then predominantly affects the motional degrees of freedom of the atoms through the dipole force. In turn, the atoms induce a phase shift on the cavity field which depends on their spatial position within the cavity mode. This regime has been investigated both with single atoms strongly coupled to optical cavities \cite{hood2000,pinkse2000} and with cold,ultracold and condensed ensembles of atoms collectively coupled to large volume cavities \cite{nagorny2003,kruse2003,black2003,klinner2006,slama2007}.
\begin{figure}
\includegraphics[width=\columnwidth]{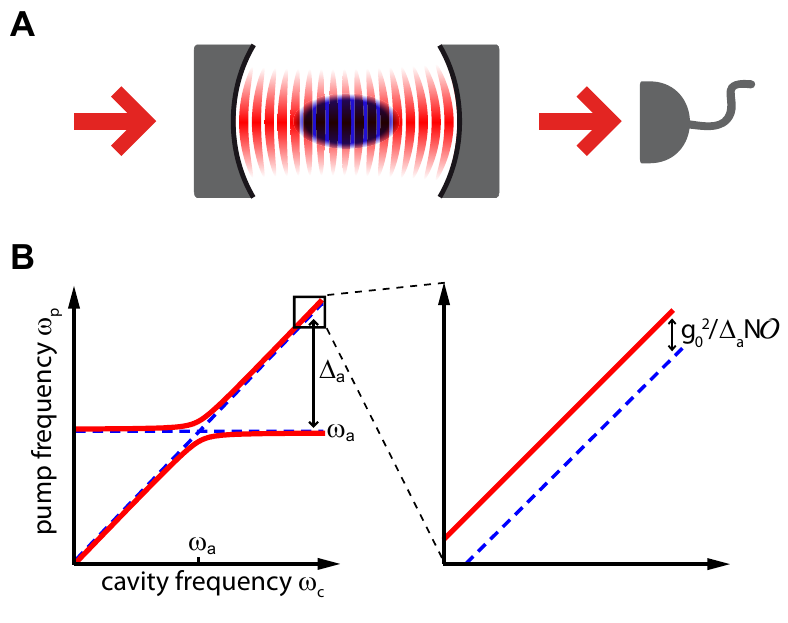}
\caption{\textbf{A.} Sketch of the coupled BEC-cavity system. To study the nonlinear coupling between condensate and
cavity light field the cavity mode is coherently driven by a pump laser while the transmitted light is monitored on a
single photon counter. \textbf{B.} Simplified energy diagram of the coupled system. We work in the dispersive regime
where the pump laser frequency $\omega_p$ is far detuned from the atomic transition frequency $\omega_a$. The collective
coupling between BEC and cavity mode leads to dressed states (solid). In the dispersive regime their energy is shifted
with respect to the bare state energies (dashed) by an amount which depends on the spatial overlap $\mathcal{O}$ between
the cavity mode and the atomic density distribution. Correspondingly, the condensate is subject to a dynamical lattice
potential whose depth depends non-locally on the atomic density distribution.} \label{figure:BECCavitySchematics}
\end{figure}

Access to a new regime has recently been attained by combining small volume ultra-high finesse optical cavities with
ultracold atomic ensembles \cite{gupta2007,murch2008} and Bose-Einstein condensates (BEC)
\cite{brennecke2007,colombe2007,brennecke2008}. Here a very strong coupling to the ensemble is achieved and the light forces
significantly influence the motion of the atoms already at the single photon level. In turn, the atoms collectively act
as a dynamical index of refraction shifting the cavity resonance according to their density distribution. Atomic motion
thus acts back on the intracavity light intensity, providing a link to cavity optomechanics \cite{hohberger2004a,schliesser2006,arcizet2006,gigan2006,corbitt2007,thompson2008,kleckner2006,lahaye2004a,teufel2008,kippenberg2007}.
Recently, bistability at photon numbers below unity \cite{gupta2007}, measurement back-action \cite{murch2008}, and triggered coherent excitations of mechanical motion \cite{brennecke2008} have been observed.

Here, we further investigate the steady state and non-steady state aspects of this highly nonlinear regime including bistability and coherent oscillations. We present experimental observations and compare them with ab-initio calculations in a mean-field approximation.

\section{Experimental setup}
\label{section_setup}

In our setup a $^{87}\mathrm{Rb}$ BEC is coupled dispersively to an ultra-high finesse Fabry-Perot optical cavity \cite{brennecke2007} (Fig.~\ref{figure:BECCavitySchematics}A). The atoms are trapped inside the cavity within a crossed beam dipole trap formed by two far-detuned laser beams oriented perpendicularly to the cavity axis. The trapping frequencies are $(\omega_x,\omega_y, \omega_z) = 2 \pi \times (220, 48, 202)$\,Hz where $x$ denotes the cavity axis and $z$ the vertical axis. The BEC contains typically $N=10^5$ atoms, which corresponds to Thomas-Fermi radii of $(R_x, R_y, R_z)$= (3.2, 19.3, 3.4) \textmu m.

The atoms are prepared in the sub-level $|F,m_F\rangle = |1, -1\rangle$ of the $5S_{1/2}$ ground state manifold, where $F$ denotes the total angular momentum and $m_F$ the magnetic quantum number. The atomic D$_2$ transition couples to a $\mathrm{TEM}_{00}$ mode of the cavity with bare frequency $\omega_c$ corresponding to a wavelength of $\lambda = 780\,\mathrm{nm}$. The mode has a waist radius of 25 \textmu m and is coherently driven at amplitude $\eta$ through one of the cavity mirrors with a circularly polarized pump laser at frequency $\omega_p$. For the experiments reported here, the pump laser was blue-detuned by $2\pi\times 58\,\mathrm{GHz}$ from the atomic transition frequency $\omega_a$. Due to a weak magnetic field oriented along the cavity axis pump photons couple only to $\sigma^+$ transitions. Summing over all accessible hyperfine levels we obtain a maximum coupling strength between a single atom and a $\sigma^+$ polarized intracavity photon of $g_0 = 2\pi \times 14.1$\,MHz. This is larger than the cavity decay rate $\kappa = 2\pi \times 1.3$\,MHz and the atomic spontaneous emission rate $\gamma = 2\pi \times 3.0$\,MHz. Therefore the condition of strong coupling is fulfilled even at the single atom level.

A piezo actuator between the cavity mirrors allows us to actively stabilize the cavity length ($\approx 178\,$\textmu m) via a Pound-Drever-Hall lock onto a far detuned laser at $829$\,nm that is tuned to resonance with a different longitudinal cavity mode. The corresponding weak standing wave potential inside the cavity has no significant influence on the results presented here.

\section{Theoretical description}
\label{section:theory}
\begin{figure}[b!]
\includegraphics[width=\columnwidth]{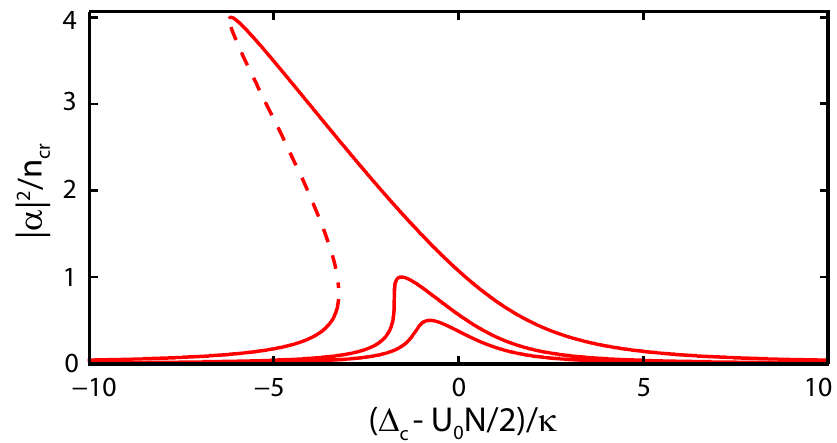}
\caption{Mean intracavity photon number $|\alpha|^2$ of the pumped BEC-cavity system versus the cavity-pump detuning $\Delta_c$ calculated for three different pump strengths $\eta = (0.7,1,2)\eta_\mathrm{cr}$ (bottom to top curve). The blue-detuned cavity light field pushes the atoms to regions of lower coupling strength which gives rise to bistability. The initially symmetric resonance curve centered around $\Delta_c=U_0 N/2$ develops above a critical pump strength $\eta_\mathrm{cr}$ a bistable region with two stable (solid lines) and one unstable branch (dashed).}\label{figure:bistability}
\end{figure}
Since the detuning $\Delta_a=\omega_p-\omega_a$ between pump laser frequency and atomic transition frequency is large compared to the collective coupling strength $\sqrt{N} g_0$ and the spontaneous emission rate $\gamma$, the population of the atomic excited state is small \cite{doherty1998}. This allows us to neglect spontaneous emission, and to eliminate the atomic excited state \cite{maschler2008} from the Tavis-Cummings Hamiltonian which describes the collective coupling between $N$ atoms and the cavity field \cite{tavis1968}. The coupling induces a light shift of the atomic ground state energy and a collective phase shift of the cavity light field (Fig.\,\ref{figure:BECCavitySchematics}B). This results in a one-dimensional optical lattice potential $\hbar U_0 \cos^2(k x)$ with the atom-light coupling varying along the cavity axis as $g_0 \cos(k x)$. Here, $U_0 = g_0^2/\Delta_a$ denotes the light shift of a maximally coupled atom in the presence of a single cavity photon, with $k = 2\pi/\lambda$. For our parameters this lattice depth is comparable to the recoil energy $\hbar \omega_\mathrm{rec} = \hbar^2 k^2/(2m)$, i.e. already mean intracavity photon numbers on the order of one are able to significantly modify the atomic density distribution. In turn, the intracavity light intensity of the driven system itself depends on the spatial distribution of the atoms in the cavity mode. The overall frequency shift of the cavity resonance is determined by the spatial overlap $\mathcal{O}$ between atomic density and cavity mode profile. Correspondingly, the coupled BEC-cavity system is governed by a strong back-action mechanism between the atomic external degree of freedom and the cavity light field.

To describe the BEC-cavity dynamics quantitatively we use a one-dimensional mean field approach. Light forces of the cavity field affecting the transverse degrees of freedom can be neglected for low intracavity photon numbers. With $\psi$ denoting the condensate wave function along the cavity axis (normalized to unity) and $\alpha$ the coherent state amplitude of the cavity field, the equations of motion read \cite{horak2000}
\begin{eqnarray}
i \hbar \dot{\psi}(x,t) &=& \Big( \frac{-\hbar^2}{2 m}\frac{\partial^2}{\partial x^2} + |\alpha(t)|^2 \hbar U_0 \cos^2(k x) \notag \\
&&+ V_\mathrm{ext}(x) + g_\mathrm{1D}|\psi|^2 \Big)\psi(x,t) \label{equation:nonlinear coupled system 1}\\
i \dot{\alpha}(t) &=& -(\Delta_c - U_0 N \mathcal{O} +i\kappa)\alpha(t) +i \eta \label{equation:nonlinear coupled system 2}.
\end{eqnarray}
Here, $V_\mathrm{ext}$ denotes the weak external trapping potential, $N$ is the total number of atoms, $g_\mathrm{1D}$ the atom-atom interaction strength integrated along the transverse directions, and $\Delta_c = \omega_p - \omega_c$ denotes the cavity-pump detuning.

These coupled equations of motion reflect that the depth of the cavity lattice potential, which is experienced by the atoms, depends non-locally on the atomic state $\psi$ via the overlap $\mathcal{O} = \langle \psi |\cos^2(kx)|\psi \rangle$. To get insight into the steady-state behavior of the condensate in this dynamical lattice potential we first solve Eq.\,\eqref{equation:nonlinear coupled system 1} for the lowest energy state in case of a fixed lattice depth. Starting from the variational ansatz
\begin{equation}\label{equation:state_expansion}
\psi(x) = c_0 + c_2 \sqrt{2} \cos(2 k x)
\end{equation}
which is appropriate for moderate lattice depths, we find the overlap integral in the ground state to be $\mathcal{O} = \frac{1}{2}-\frac{|\alpha|^2 U_0}{16 \omega_\mathrm{rec}}$. Here, the external trapping potential $V_\mathrm{ext}$ and atom-atom interactions have been neglected for simplicity. Correspondingly, the BEC acts as a Kerr medium that shifts the empty cavity resonance proportionally to the intracavity light intensity. After inserting this result into the steady state solution of Eq.~\eqref{equation:nonlinear coupled system 2}
\begin{equation}\label{equation:consistency}
|\alpha|^2 = \frac{\eta^2}{\kappa^2+(\Delta_c-U_0 N \mathcal{O})^2},
\end{equation}
an algebraic equation of third order in $|\alpha|^2$ is obtained which determines the resonance curve of the system. For sufficient pump strength $\eta$ the system exhibits bistable behavior (Fig.\,\ref{figure:bistability}), a property which is known from optical and mechanical Kerr nonlinearity \cite{boyd2008,dorsel1983,gozzini1985,meystre1985,nagorny2003,gupta2007}. Namely, while increasing the pump strength $\eta$ the initially Lorentzian resonance curve of height $\eta^2/\kappa^2$ gets asymmetric and develops an increasing region with three possible steady states above a critical value $\eta_\mathrm{cr}$. A detailed analysis results in a corresponding critical intracavity photon number on resonance of $n_\mathrm{cr} = \frac{8}{3 \sqrt{3}}\frac{16 \kappa \omega_\mathrm{rec}}{N U_0^2}$.

\section{Bistability measurement}
\begin{figure}[t!]
\includegraphics[width=\columnwidth]{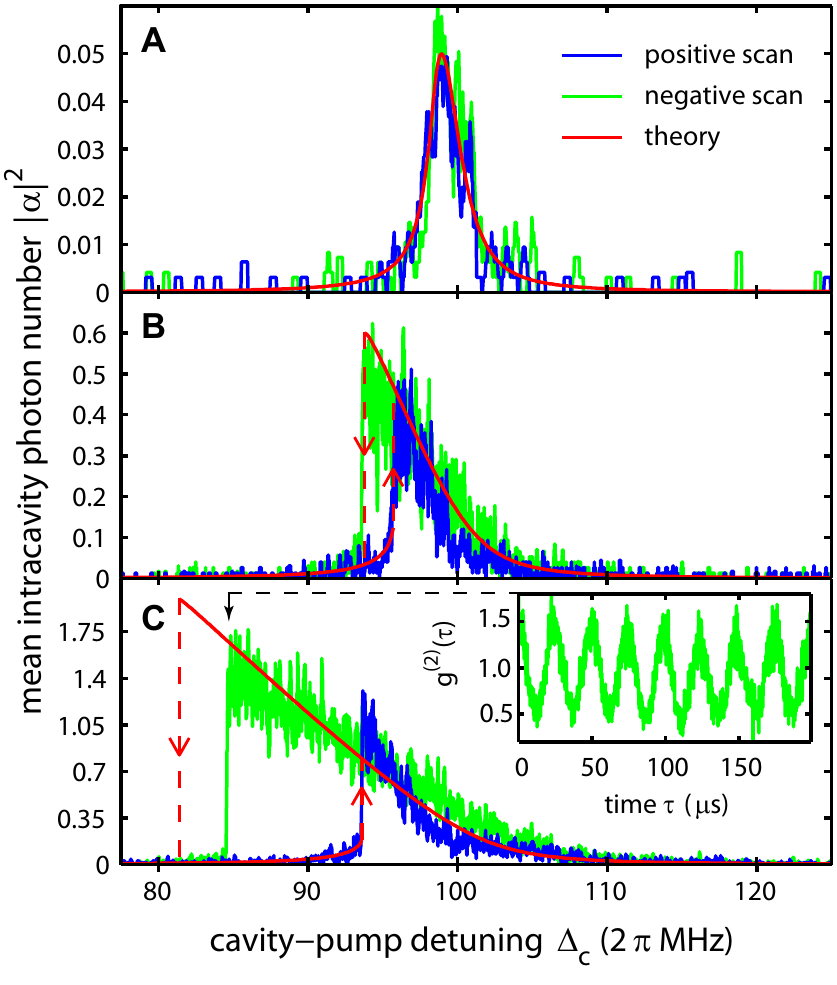}
\caption{Bistable behavior at low photon number. The traces show the mean intracavity photon number $|\alpha|^2$ versus the cavity-pump detuning $\Delta_c$. Traces A, B and C correspond to pump strengths of $\eta = (0.22,0.78,1.51)\kappa$, respectively. The intracavity photon number is deduced from the detector count rate. Each graph corresponds to a single experimental sequence during which the pump laser frequency was scanned twice across the resonance, first with increasing detuning $\Delta_c$ (blue curve) and then with decreasing detuning (green curve). The scan speed was $2\pi\times 1\,\mathrm{MHz/ms}$ and the raw data has been averaged over 400\,\textmu s (A) and 100\,\textmu s (B and C). We corrected for a drift of the resonance caused by a measured atom loss rate of $92/\mathrm{ms}$ assumed to be constant during the measurement. The theoretically expected stable resonance branches (red) have been calculated for $10^5$ atoms (deduced from absorption images) taking a transverse part of the mode overlap of 0.6 into account. This value was deduced from several scans across the resonance in the non-bistable regime and is about 25\% below the value expected from the BEC and cavity mode geometry. Shot-to-shot fluctuations in the atom number resulting in uncontrolled frequency shifts were corrected for by overlapping the individual data traces A,B and C with the theoretically expected curves. The inset of C shows photon-photon correlations of the green trace calculated from the last 400\,\textmu s right before the system transits to the lower stable branch. Due to averaging these oscillations are not visible in the main graph.} \label{figure:scan over resonance}
\end{figure}
\begin{figure*}[t!]
\includegraphics{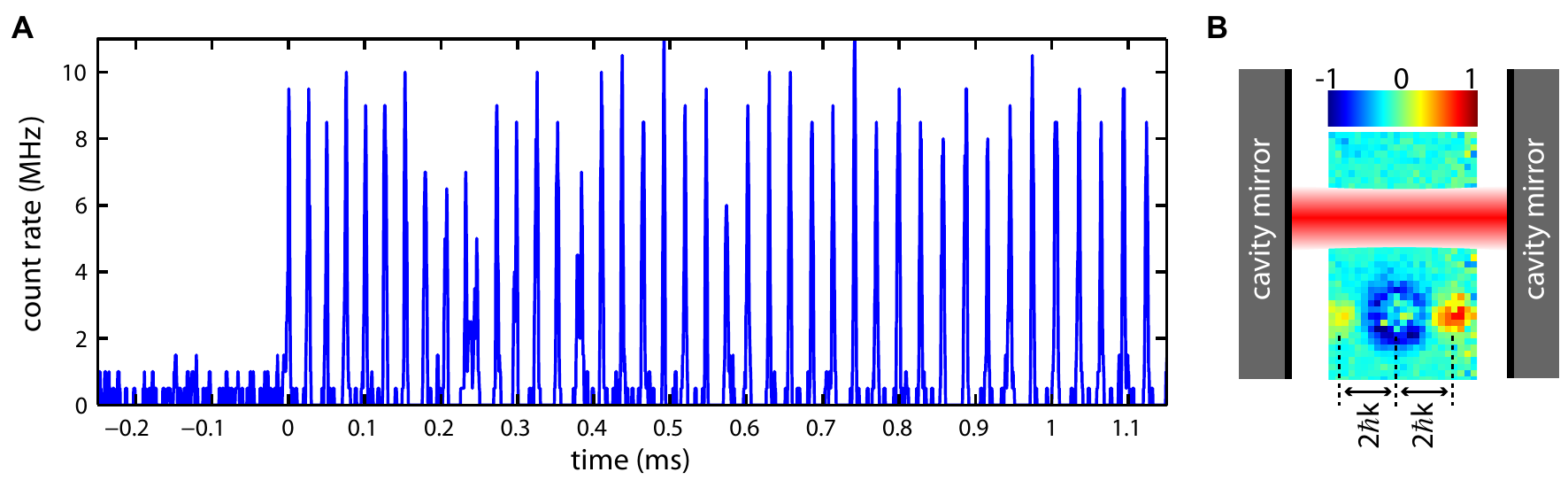}
\caption{\textbf{A.} Coherent dynamics of the BEC in the dynamical lattice potential. Shown is the count rate of the single photon detector while scanning with increasing cavity-pump detuning across the bistable resonance curve. The scan speed was set to $2\pi\times 2\,\mathrm{MHz/ms}$ with a maximum intracavity photon number of 9.5. The condensate is excited due to the non-adiabatic branch transition resulting in oscillations of the overlap $\mathcal{O}$ clearly visible in a periodic cavity output.  \textbf{B.} Absorption image revealing the population in the $|p = \pm2\hbar k\rangle$ momentum components during the coherent oscillations. Once the coherent dynamics was excited both trapping potential and pump laser were switched off and the cloud was imaged after 4\,ms free expansion. To clearly detect the small $|p = \pm2\hbar k\rangle$ population we averaged over 9 independent images and subtracted the average of 9 different images without excitation (taken after the oscillations had stopped \cite{brennecke2008}).} \label{figure:oscillations}
\end{figure*}
To study the nonlinear coupling between BEC and cavity field experimentally, the pump laser frequency was scanned slowly (compared to the atomic motion) across the resonance while recording the cavity transmission on a single photon counter. From the measured photon count rate the mean intracavity photon number is deduced by correcting for the quantum efficiency ($\approx 0.5$) and the saturation of the single photon counter, and by taking into account the transmission (2.3\,ppm) of the output coupling mirror as well as the losses at the detection optics (15\,\%). The systematic uncertainty in determining the intracavity photon number is estimated to be 25\,\%.

Typical resonance curves obtained for different pump strengths are shown in Fig.\,\ref{figure:scan over resonance}. For maximum intracavity photon numbers well below the critical photon number $n_\mathrm{cr}$ the resonance curve is Lorentzian shaped and does not depend on the scan direction of the pump laser (A). When increasing the pump strength beyond the critical value we observe a pronounced asymmetry of the resonance and hysteretic behavior which indicates bistability of the system (B). The frequency range over which bistability occurs gets enlarged by further increasing the pump strength (C).

We compare our experimental data with resonance curves obtained from a numerical solution of the coupled set of Eqs.~\eqref{equation:nonlinear coupled system 1} and \eqref{equation:nonlinear coupled system 2} including atom-atom interactions and the external trapping potential (red lines in Fig.\,\ref{figure:scan over resonance}). We find a critical photon number of $n_\mathrm{cr} = 0.21$, in accordance with our experimental observations within the systematic uncertainties. The inclusion of atom-atom interactions results in a critical photon number which is slightly larger than the value 0.18 obtained from the analytical interaction-free model. For very low photon numbers (A and B) we find good agreement between the measured and calculated resonance curves. However, for increasing pump strengths we observe that the system deviates more and more from the calculated steady state curves (C). This is visible in a precipitate transition from the upper branch to the lower one while scanning with decreasing $\Delta_c$. Such deviations indicate a superposed non-steady state dynamics. This dynamics is governed by the inertia of our refractive index medium, and goes beyond the physics of a pure Kerr medium \cite{fabre1994}. Experimentally, this is supported by detecting regular oscillations in the second order correlation function $g^{(2)}(\tau)$ which was evaluated from the transmission signal right before the system leaves the upper resonance branch (Fig.~\ref{figure:scan over resonance}C inset).

\section{Dynamics}

Coherent non-steady state dynamics of the system can also be excited more directly by means of a non-adiabatic increase in the cavity light intensity. This is naturally provided by the sudden transition which appears while scanning with increasing $\Delta_c$ across the bistable resonance (Fig.~\ref{figure:oscillations}A). Once the system reaches the turning point of the lower stable branch (Fig.~\ref{figure:bistability}) a periodic dynamics is excited which gets observable through a strongly pulsed cavity transmission. This dynamics has been reported on previously \cite{brennecke2008}. In short, a small fraction of condensate atoms is scattered by the cavity lattice into the higher momentum states $|p = \pm 2\hbar k\rangle$. Due to matter-wave interference with the remaining $|p=0\rangle$ atoms the atomic cloud develops a density oscillation which shifts the system periodically in resonance with the pump laser. Direct evidence for the coherence of this dynamics is obtained by recording the atomic momentum distribution via absorption imaging (Fig.~\ref{figure:oscillations}B).

Further insight into the non-steady state behavior can be gained from the analogy between the coupled BEC-cavity system
and a mechanical oscillator coupled to a cavity field via radiation pressure \cite{brennecke2008}. Our mechanical
oscillator can be identified with the $c_2$-mode in the state expansion Eq.\,\eqref{equation:state_expansion}.
Matter-wave interference with the $c_0$-mode gives rise to a spatial modulation of the atomic density, and results in
harmonic oscillations of the overlap $\mathcal{O}$ at a frequency of $4 \omega_{\mathrm{rec}} \approx 2 \pi \times
15\,\mathrm{kHz}$.

This mapping to cavity optomechanics shown in \cite{brennecke2008} helps to gain knowledge on the dynamical behavior of the system. Since the mechanical oscillator is subject to the radiation pressure force its stiffness is modified according to the intracavity light intensity. This mechanism is known in the literature as 'optical spring' \cite{dorsel1983,sheard2004}. We observe a clear signature of this effect in the photon-photon correlations (Fig.~\ref{figure:scan over resonance}C inset) oscillating at approximately 42\,kHz which is a factor of 2.9 larger than the bare oscillator frequency. A detailed study of this dynamics including the amplification effects due to retardation between cavity light field and oscillator motion is the subject of ongoing work.

\section{Conclusion}

Here we have studied the dynamical coupling between a BEC and a cavity optical lattice. We have observed a strong optical nonlinearity at the single photon level, manifested by bistable behavior and coherent oscillations around the steady state. These results complement the cavity optomechanical studies traditionally conducted on microfabricated or high precision interferometric devices (for a recent review see \cite{kippenberg2008}). Our system has remarkable properties which should allow us to experimentally explore the quantum regime of cavity optomechanics \cite{kippenberg2008,ludwig2008}. The mechanical oscillator intrinsically starts in the ground state, from which, due to collective enhancement of the coupling, a single motional excitation can cause a shift of the cavity resonance on the order of the cavity linewidth. Inversely, a change of one photon in the light field strongly modifies the atomic motional state. Beyond the classical nonlinear observations reported here, the system is therefore promising to reveal signatures of the quantum nature of the light and matter fields \cite{murch2008,ludwig2008,rai2008}.

\begin{acknowledgements}
We acknowledge stimulating discussions with I. Carusotto and financial support from QSIT and the EU-programs SCALA and OLAQUI. C. G. acknowledges ETH fellowship support.
\end{acknowledgements}


\bibliographystyle{ApplPhysB}

\begin{thebibliography}{10}
\providecommand{\url}[1]{\texttt{#1}}
\providecommand{\urlprefix}{URL }
\providecommand{\eprint}[2][]{\url{#2}}

\bibitem{haroche2006}
S.~{H}aroche, J.~M. {R}aimond, \emph{{E}xploring the {Q}uantum} (Oxford
  University Press, Oxford, 2006)

\bibitem{raimond2001}
J.~M. {R}aimond, M.~{B}rune, S.~{H}aroche, {R}ev. {M}od. {P}hys. \textbf{73},
  565 (2001)

\bibitem{walther2002}
H.~{W}alther, {A}dv. {C}hem. {P}hys. \textbf{122}, 167 (2002)

\bibitem{kimble1998}
H.~J. {K}imble, {P}hysica {S}cripta \textbf{T76}, 127 (1998)

\bibitem{maunz2005a}
P.~{M}aunz, T.~{P}uppe, I.~{S}chuster, N.~{S}yassen, P.~W.~H. {P}inkse,
  G.~{R}empe, {P}hys. {R}ev. {L}ett. \textbf{94}, 033002 (2005)

\bibitem{mundt2002}
A.~B. {M}undt, A.~{K}reuter, C.~{B}echer, D.~{L}eibfried, J.~{E}schner,
  F.~{S}chmidt {K}aler, R.~{B}latt, {P}hys. {R}ev. {L}ett. \textbf{89}, 103001
  (2002)

\bibitem{wallraff2004}
A.~{W}allraff, D.~I. {S}chuster, A.~{B}lais, L.~{F}runzio, R.-S. {H}uang,
  J.~{M}ajer, S.~{K}umar, S.~M. {G}irvin, R.~J. {S}choelkopf, {N}ature
  \textbf{431}, 162 (2004)

\bibitem{hood2000}
C.~J. {H}ood, T.~W. {L}ynn, A.~C. {D}oherty, A.~S. {P}arkins, H.~J. {K}imble,
  {S}cience \textbf{287}, 1447 (2000)

\bibitem{pinkse2000}
P.~W.~H. {P}inkse, T.~{F}ischer, P.~{M}aunz, G.~{R}empe, {N}ature \textbf{404},
  365 (2000)

\bibitem{nagorny2003}
B.~{N}agorny, {T}h. {E}ls\"asser, A.~{H}emmerich, {P}hys. {R}ev. {L}ett.
  \textbf{91}, 153003 (2003)

\bibitem{kruse2003}
D.~{K}ruse, C.~von {C}ube, C.~{Z}immermann, P.~W. {C}ourteille, {P}hys. {R}ev.
  {L}ett. \textbf{91}, 183601 (2003)

\bibitem{black2003}
A.~T. {B}lack, H.~W. {C}han, V.~{V}uleti\'{c}, {P}hys. {R}ev. {L}ett.
  \textbf{91}, 203001 (2003)

\bibitem{klinner2006}
J.~{K}linner, M.~{L}indholdt, B.~{N}agorny, A.~{H}emmerich, {P}hys. {R}ev.
  {L}ett. \textbf{96}, 023002 (2006)

\bibitem{slama2007}
S.~{S}lama, S.~{B}ux, G.~{K}renz, C.~{Z}immermann, P.~W. {C}ourteille, {P}hys.
  {R}ev. {L}ett. \textbf{98}, 053603 (2007)

\bibitem{gupta2007}
S.~{G}upta, K.~L. {M}oore, K.~W. {M}urch, D.~M. {S}tamper {K}urn, {P}hys.
  {R}ev. {L}ett. \textbf{99}, 213601 (2007)

\bibitem{murch2008}
K.~W. {M}urch, K.~L. {M}oore, S.~{G}upta, D.~M. {S}tamper {K}urn, {N}ature
  {P}hysics \textbf{4}, 561 (2008)

\bibitem{brennecke2007}
F.~{B}rennecke, T.~{D}onner, S.~{R}itter, T.~{B}ourdel, M.~{K}\"ohl,
  T.~{E}sslinger, {N}ature \textbf{450}, 268 (2007)

\bibitem{colombe2007}
Y.~{C}olombe, T.~{S}teinmetz, G.~{D}ubois, F.~{L}inke, D.~{H}unger,
  J.~{R}eichel, {N}ature \textbf{450}, 272 (2007)

\bibitem{brennecke2008}
F.~{B}rennecke, S.~{R}itter, T.~{D}onner, T.~{E}sslinger, {S}cience
  \textbf{322}, 235 (2008)

\bibitem{hohberger2004a}
C.~{H}\"ohberger {M}etzger, K.~{K}arrai, {N}ature \textbf{432}, 1002 (2004)

\bibitem{schliesser2006}
A.~{S}chliesser, P.~{D}el'{H}aye, N.~{N}ooshi, K.~J. {V}ahala, T.~J.
  {K}ippenberg, {P}hys. {R}ev. {L}ett. \textbf{97}, 243905 (2006)

\bibitem{arcizet2006}
O.~{A}rcizet, P.-F. {C}ohadon, T.~{B}riant, M.~{P}inard, A.~{H}eidmann,
  {N}ature \textbf{444}, 71 (2006)

\bibitem{gigan2006}
S.~{G}igan, H.~R. {B}\"ohm, M.~{P}aternostro, F.~{B}laser, G.~{L}anger, J.~B.
  {H}ertzberg, K.~C. {S}chwab, D.~{B}\"auerle, M.~{A}spelmeyer, A.~{Z}eilinger,
  {N}ature \textbf{444}, 67 (2006)

\bibitem{corbitt2007}
T.~{C}orbitt, Y.~{C}hen, E.~{I}nnerhofer, H.~{M}\"uller {E}bhardt,
  D.~{O}ttaway, H.~{R}ehbein, D.~{S}igg, S.~{W}hitcomb, C.~{W}ipf,
  N.~{M}avalvala, {P}hys. {R}ev. {L}ett. \textbf{98}, 150802 (2007)

\bibitem{thompson2008}
J.~D. {T}hompson, B.~M. {Z}wickl, A.~M. {J}ayich, F.~{M}arquardt, S.~M.
  {G}irvin, J.~G.~E. {H}arris, {N}ature \textbf{452}, 72 (2008)

\bibitem{kleckner2006}
D.~{K}leckner, D.~{B}ouwmeester, {N}ature \textbf{444}, 75 (2006)

\bibitem{lahaye2004a}
M.~D. {L}a{H}aye, O.~{B}uu, B.~{C}amarota, K.~C. {S}chwab, {S}cience
  \textbf{304}, 74 (2004)

\bibitem{teufel2008}
J.~D. {T}eufel, J.~W. {H}arlow, C.~A. {R}egal, K.~W. {L}ehnert, {P}hys. {R}ev. {L}ett. \textbf{101}, 197203
  (2008)

\bibitem{kippenberg2007}
T.~J. {K}ippenberg, K.~J. {V}ahala, {O}pt. {E}xpress \textbf{15}, 17172 (2007)

\bibitem{doherty1998}
A.~C. {D}oherty, A.~S. {P}arkins, S.~M. {T}an, D.~F. {W}alls, {P}hys. {R}ev.
  {A} \textbf{57}, 4804 (1998)

\bibitem{maschler2008}
C.~{M}aschler, I.~B. {M}ekhov, H.~{R}itsch, {T}he {E}uropean {P}hysical
  {J}ournal {D} \textbf{46}, 545 (2008)

\bibitem{tavis1968}
M.~{T}avis, F.~W. {C}ummings, {P}hys. {R}ev. \textbf{170}, 379 (1968)

\bibitem{horak2000}
P.~{H}orak, S.~M. {B}arnett, H.~{R}itsch, {P}hys. {R}ev. {A} \textbf{61},
  033609 (2000)

\bibitem{boyd2008}
R.~W. {B}oyd, \emph{{N}onlinear {O}ptics} (Academic Press, Boston, 2008)

\bibitem{dorsel1983}
A.~{D}orsel, J.~D. {M}c{C}ullen, P.~{M}eystre, E.~{V}ignes, H.~{W}alther,
  {P}hys. {R}ev. {L}ett. \textbf{51}, 1550 (1983)

\bibitem{gozzini1985}
A.~{G}ozzini, F.~{M}accarrone, F.~{M}ango, I.~{L}ongo, S.~{B}arbarino, {J}.
  {O}pt. {S}oc. {A}m. {B} \textbf{2}, 1841 (1985)

\bibitem{meystre1985}
P.~{M}eystre, E.~M. {W}right, J.~D. {M}c{C}ullen, E.~{V}ignes, {J}. {O}pt.
  {S}oc. {A}m. {B} \textbf{2}, 1830 (1985)

\bibitem{fabre1994}
C.~{F}abre, M.~{P}inard, S.~{B}ourzeix, A.~{H}eidmann, E.~{G}iacobino,
  S.~{R}eynaud, {P}hys. {R}ev. {A} \textbf{49}, 1337 (1994)

\bibitem{sheard2004}
B.~S. {S}heard, M.~B. {G}ray, C.~M. {M}ow {L}owry, D.~E. {M}c{C}lelland, S.~E.
  {W}hitcomb, {P}hys. {R}ev. {A} \textbf{69}, 051801 (2004)

\bibitem{kippenberg2008}
T.~J. {K}ippenberg, K.~J. {V}ahala, {S}cience \textbf{321}, 1172 (2008)

\bibitem{ludwig2008}
M.~{L}udwig, B.~{K}ubala, F.~{M}arquardt, {N}ew {J}. {P}hys.  \textbf{10}, 095013 (2008).

\bibitem{rai2008}
A.~{R}ai, G.~S. {A}garwal, {P}hys. {R}ev. {A} \textbf{78}, 013831 (2008)

\end{thebibliography}

\end{document}